\documentclass[twocolumn]{aastex631}

\received{\today}
\revised{\today}
\accepted{\today}

\submitjournal{ApJL}

\shorttitle{Discovery of the Longest-Period Classical Cepheid in the Milky Way}
\shortauthors{Soszy\'nski et al.}

\begin{document}

\title{Discovery of the Longest-Period Classical Cepheid in the Milky Way}

\author[0000-0002-7777-0842]{I. Soszy\'nski}
\affiliation{Astronomical Observatory, University of Warsaw, Al.~Ujazdowskie~4, 00-478~Warszawa, Poland}
\email{soszynsk@astrouw.edu.pl}

\author[0000-0001-9439-604X]{D. M. Skowron}
\affiliation{Astronomical Observatory, University of Warsaw, Al.~Ujazdowskie~4, 00-478~Warszawa, Poland}

\author[0000-0001-5207-5619]{A. Udalski}
\affiliation{Astronomical Observatory, University of Warsaw, Al.~Ujazdowskie~4, 00-478~Warszawa, Poland}

\author[0000-0002-2339-5899]{P. Pietrukowicz}
\affiliation{Astronomical Observatory, University of Warsaw, Al.~Ujazdowskie~4, 00-478~Warszawa, Poland}

\author[0000-0002-1650-1518]{M. Gromadzki}
\affiliation{Astronomical Observatory, University of Warsaw, Al.~Ujazdowskie~4, 00-478~Warszawa, Poland}

\author[0000-0002-0548-8995]{M. K. Szyma\'nski}
\affiliation{Astronomical Observatory, University of Warsaw, Al.~Ujazdowskie~4, 00-478~Warszawa, Poland}

\author[0000-0002-2335-1730]{J. Skowron}
\affiliation{Astronomical Observatory, University of Warsaw, Al.~Ujazdowskie~4, 00-478~Warszawa, Poland}

\author[0000-0001-7016-1692]{P. Mr\'oz}
\affiliation{Astronomical Observatory, University of Warsaw, Al.~Ujazdowskie~4, 00-478~Warszawa, Poland}

\author[0000-0002-9245-6368]{R. Poleski}
\affiliation{Astronomical Observatory, University of Warsaw, Al.~Ujazdowskie~4, 00-478~Warszawa, Poland}

\author[0000-0003-4084-880X]{S. Koz\l{}owski}
\affiliation{Astronomical Observatory, University of Warsaw, Al.~Ujazdowskie~4, 00-478~Warszawa, Poland}

\author[0000-0002-6212-7221]{P. Iwanek}
\affiliation{Astronomical Observatory, University of Warsaw, Al.~Ujazdowskie~4, 00-478~Warszawa, Poland}

\author[0000-0002-3051-274X]{M. Wrona}
\affiliation{Astronomical Observatory, University of Warsaw, Al.~Ujazdowskie~4, 00-478~Warszawa, Poland}

\author[0000-0001-6364-408X]{K. Ulaczyk}
\affiliation{Department of Physics, University of Warwick, Gibbet Hill Road, Coventry, CV4~7AL,~UK}
\affiliation{Astronomical Observatory, University of Warsaw, Al.~Ujazdowskie~4, 00-478~Warszawa, Poland}

\author[0000-0002-9326-9329]{K. Rybicki}
\affiliation{Department of Particle Physics and Astrophysics, Weizmann Institute of Science, Rehovot 76100, Israel}
\affiliation{Astronomical Observatory, University of Warsaw, Al.~Ujazdowskie~4, 00-478~Warszawa, Poland}

\author[0000-0002-8911-6581]{M. Mr\'oz}
\affiliation{Astronomical Observatory, University of Warsaw, Al.~Ujazdowskie~4, 00-478~Warszawa, Poland}

\begin{abstract}

We report the discovery of the classical Cepheid OGLE-GD-CEP-1884 (=~GDS\_J1535467-555656) with the longest pulsation period known in our Galaxy. The period of 78.14~d is nearly 10~d longer than that of the previous record-holding Cepheid, S Vulpeculae, and thus, OGLE-GD-CEP-1884 can be categorized as the first ultra long period Cepheid in the Milky Way. This star is present in the ASAS-SN and {\it Gaia} DR3 catalogs of variable stars, but it has been classified as a long-period variable in those catalogs. Based on more than 10 years of the photometric monitoring of this star carried out by the OGLE project in the {\it I} and {\it V} bands and a radial velocity curve from the {\it Gaia} Focused Product Release, we unequivocally demonstrate that this object is a fundamental-mode classical Cepheid. By employing the mid-infrared period--luminosity relation, we determine the distance to OGLE-GD-CEP-1884 ($4.47\pm0.34$~kpc) and place it on the Milky Way map, along with about 2400 other classical Cepheids. We also discuss the potential of finding additional ultra long period Cepheids in our Galaxy.

\end{abstract}

\keywords{stars: variables: Cepheids --- stars: oscillations --- stars: distances}

\section{Introduction}\label{sec:intro}

The Optical Gravitational Lensing Experiment (OGLE) is a long-term sky survey involving regular photometric observations of about 2~billion stars in our Galaxy and in the Magellanic Clouds \citep{2015AcA....65....1U}. One of the major achievements of the project is the OGLE Collection of Variable Stars (OCVS) -- a catalog currently comprising approximately 1.1~million variable stars of various types. Compared to other modern large catalogs of variable stars, OCVS distinguishes itself through the method used for the object classification, which relies on the visual inspection of light curves. Thanks to this traditional approach, the OGLE group has identified several new classes and subclasses of variable stars, e.g. peculiar W~Virginis stars \citep{2008AcA....58..293S}, anomalous double-mode RR~Lyrae stars \citep{2016MNRAS.463.1332S}, or Blue Large-Amplitude Pulsators \citep[BLAPs;][]{2017NatAs...1E.166P}.

The extensive OGLE databases have been utilized to discover about half of the classical Cepheids currently known in the Milky Way \citep{2018AcA....68..315U,2020AcA....70..101S,2021AcA....71..205P}. Classical Cepheids (also known as $\delta$~Cephei variables or type~I Cepheids) are relatively young (${\la}400$~Myr) and luminous ($10^2$--$10^5\,{\rm L}_\odot$) radially pulsating stars, 4--20 times more massive than the Sun. The best-known characteristic of Cepheids is the period--luminosity relation (Leavitt law) they fulfill, which is commonly employed for measuring intergalactic distances \citep[e.g.][]{2016ApJ...826...56R} and studying the structure of galaxies \citep[e.g.][]{2019Sci...365..478S,2019AcA....69..305S}. The pulsation periods of fundamental-mode Cepheids range from 1 to over 200~d, but the majority oscillate with periods below 10~d. Above this limit, the number of Cepheids decreases with increasing periods. The longest-period classical Cepheid known so far in our Galaxy was S~Vulpeculae \citep{1970AJ.....75..244F,2021AcA....71..205P}, with a period of 68.65~d.

Recently, \citet{2023A&A...680A..36G} published radial velocity curves for 9614 stars classified as long-period variables (LPVs), or, in other words, pulsating red giants. Most of these objects are relatively bright, above the saturation limit of the OGLE photometry ($I\approx11$~mag). Nevertheless, we successfully extracted OGLE light curves for over a thousand stars from this list. A visual inspection of these data confirmed that the vast majority of them are indeed LPVs. However, one variable object, with a period of 78.14~d, drew our special attention due to the distinctive morphology of its light curve.

OGLE-GD-CEP-1884 (also known as GDS\_J1535467-555656, ASASSN-V J153546.74-555656.1, {\it Gaia} DR3 5883587875302126592; hereafter OGLE-GD-CEP-1884) was first identified as a variable star (albeit without a specific classification) by \citet{2015AN....336..590H} based on photometric data acquired during the Bochum Galactic Disk Survey (GDS). Then, \citet{2019MNRAS.486.1907J} examined the light curve of this star from the All-Sky Automated Survey for Supernovae (ASAS-SN) and categorized it as a semi-regular variable -- one of the subclasses of LPVs. Finally, this object was included in the {\it Gaia} Data Release 3 (DR3) catalog of variable stars \citep{2023A&A...674A..15L}, where it was also classified as an LPV. In this paper, we demonstrate that the ASAS-SN and {\it Gaia} classifications of OGLE-GD-CEP-1884 are incorrect. The shapes of the light and radial velocity curves unequivocally indicate that it is a classical Cepheid with the longest pulsation period known in the Milky Way. Up to now, this star has been absent from the OGLE collection of classical Cepheids \citep{2018AcA....68..315U,2020AcA....70..101S} because the OGLE search for Cepheids was restricted to objects with periods shorter than 50~d. In Table~\ref{tab:table1}, we provide basic observational parameters of OGLE-GD-CEP-1884, including its equatorial and Galactic coordinates, pulsation period, mean magnitudes in various passbands, parallax, proper motion, and mean radial velocity from {\it Gaia} DR3.

\begin{deluxetable}{rl}
\tablecaption{Parameters of OGLE-GD-CEP-1884\label{tab:table1}}
\startdata
&\\
Right ascension [J2000.0]: & 15:35:46.70 \\
Declination [J2000.0]: & $-$55:56:56.4 \\
Galactic longitude [\degr]: & 324.522733 \\
Galactic latitude [\degr]: & $-$0.125534 \\
Pulsation period [d]: & $78.140\pm0.011$  \\
Mean {\it V}-band magnitude (OGLE): & 16.83 \\
Mean {\it I}-band magnitude (OGLE): & 11.40 \\
Mean {\it G}-band magnitude ({\it Gaia}): & 12.84 \\
Mean {\it J}-band magnitude (2MASS):  & ~7.47 \\
Mean {\it H}-band magnitude (2MASS):  & ~6.14 \\
Mean {\it K$_S$}-band magnitude (2MASS): & ~5.51 \\
Parallax ({\it Gaia}) [mas]: & $0.176\pm0.077$ \\
Proper motion ({\it Gaia}) [mas/yr]: & $5.88\pm0.08$ \\
Mean radial velocity ({\it Gaia}) [km/s]: & $-72.3\pm0.4$ \\
Velocity component $U$ [km/s]: & $-9.9\pm8.5$ \\
Velocity component $V$ [km/s]: & $220.8\pm3.9$ \\
Velocity component $W$ [km/s]: & $-3.4\pm1.8$ \\
\enddata
\end{deluxetable}

\section{Observational Data}\label{sec:obs}

The {\it VI}-band observations of OGLE-GD-CEP-1884 were obtained as part of the OGLE-IV survey \citep{2015AcA....65....1U} carried out with the 1.3-meter Warsaw Telescope at Las Campanas Observatory in Chile. The telescope is equipped with a mosaic camera comprising 32 CCD chips with a pixel scale of 0.26~arcsec and a total field of view of 1.4~square degrees. Around 180 {\it I}-band observations of our target were secured between May 2013 and March 2024, with a break of more than two years (2020--2022) during the COVID-19 pandemic. The integration time of the {\it I}-band observations was 30~s. In the {\it V}-band filter, a total of 26 data points were collected, including 12 measurements obtained in the years 2010-2011, 5 observations in 2015-2016, and 9 points gathered between January and March 2024. The initial {\it V}-band observations were conducted with an integration time of 180~s. Then, starting from 2015, the exposure time was reduced to 30~s. A comprehensive description of the instrumentation, photometric reductions, and astrometric calibrations of the OGLE-IV data can be found in the work of \citet{2015AcA....65....1U}.

\section{Classification of OGLE-GD-CEP-1884}\label{sec:dis}

\begin{figure}
\centering
\includegraphics[width=7cm]{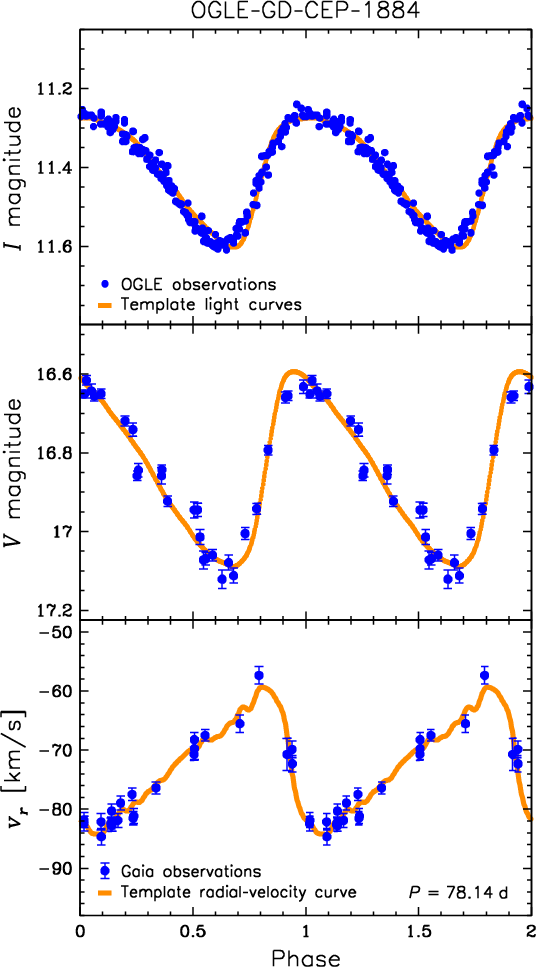}
\caption{Phase-folded {\it I}-band (upper panel) and {\it V}-band (middle panel) OGLE light curves of OGLE-GD-CEP-1884 and radial-velocity curve of this star from the {\it Gaia} Focused Product Release \citep[][lower panel]{2023A&A...680A..36G}. The orange curves depict template light and radial-velocity curves of a classical Cepheid with a period of 78.14~d generated using the code provided by \citet{2012ApJ...748..107P}.\label{fig1}}
\end{figure}

The upper and middle panels of Figure~\ref{fig1} display the {\it I} and {\it V}-band light curves of OGLE-GD-CEP-1884, folded with the period of 78.14~d. Throughout more than a decade of OGLE photometric monitoring, the period, amplitudes, and shapes of the light curves have consistently remained stable, exhibiting only slight fluctuations (up to $\pm0.04$~mag in the {\it I} band) in average brightness. These fluctuations manifest as the minor scattering of points in the light curves presented in Figure~\ref{fig1}. Such variations could be attributed to observational errors or may indicate actual changes occurring within the star or in the circumstellar and interstellar matter between us and the object.

The orange curves drawn together with the points in Figure~\ref{fig1} do not depict arbitrary functions fitted to the OGLE photometry. Instead, they represent template light curves of a classical Cepheid with a period of 78.14~d, calculated using the code provided by \citet{2012ApJ...748..107P}. Fitting these model light curves to the observations only involved scaling their amplitudes and adjusting the mean brightness and phases. A good agreement between the shapes of the observed and template light curves is visible in Figure~\ref{fig1}, strongly supporting the hypothesis that OGLE-GD-CEP-1884 is a classical Cepheid. Also, the observed amplitude ratio and relative phase shift between {\it I} and {\it V}-band light curves satisfactorily agree with the model. Small differences between the templates and observations ($\la0.05$~mag) may arise from the different metallicity of our star compared to that assumed in the model \citep{2012ApJ...748..107P}, as well as minor deviations of the OGLE-IV filters from the standard filters of the Johnson-Cousins photometric system \citep{2015AcA....65....1U}.

\begin{figure*}
\centering
\includegraphics[width=16cm]{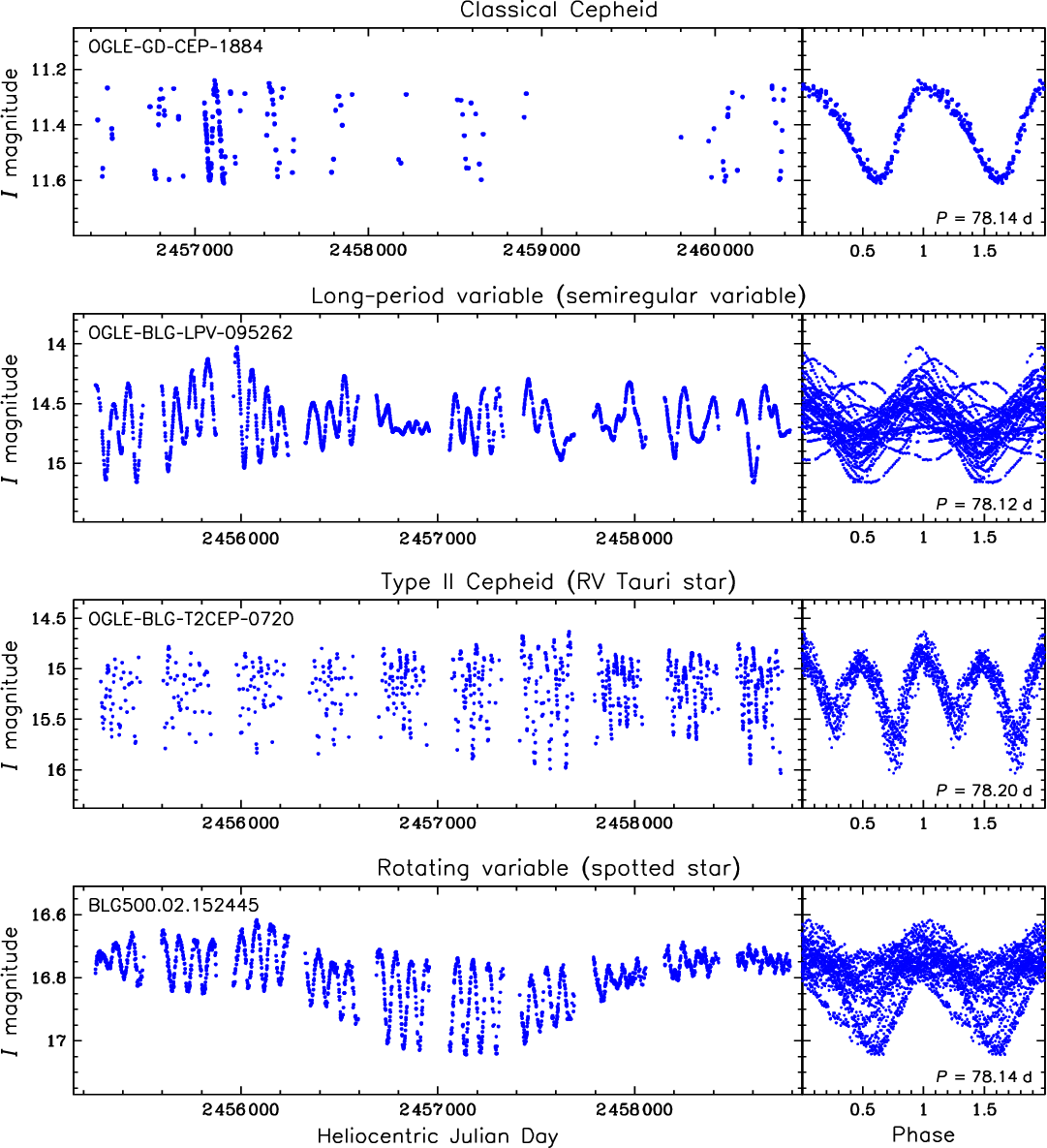}
\caption{Unfolded (left panels) and phase-folded (right panels) {\it I}-band light curves of OGLE-GD-CEP-1884 (upper panels) and three example variable stars with periods nearly identical to that of OGLE-GD-CEP-1884: a semiregular variable from the OGLE collection of LPVs in the Galactic bulge \citep{2013AcA....63...21S}, an RV Tauri star from the OGLE collection of type II Cepheids in the bulge \citep{2017AcA....67..297S}, and a spotted variable star from the Galactic bulge \citep{2019ApJ...879..114I}. Note the differences between the light curve of a classical Cepheid and those of other types of variable stars.\label{fig2}}
\end{figure*}

Let us explore the possibility that OGLE-GD-CEP-1884 may be a variable star of a different type than a classical Cepheid. This object has been categorized as an LPV in both the ASAS-SN \citep{2019MNRAS.486.1907J} and {\it Gaia} DR3 \citep{2023A&A...674A..15L} catalogs of variable stars, presumably due to its apparent red color ($V-I=5.43$~mag, $G_{\rm{BP}}-G_{\rm{RP}}=5.25$~mag). In Figure~\ref{fig2}, we once again present the {\it I}-band light curve of OGLE-GD-CEP-1884 (upper panels), but this time also in the unfolded version (left panel). Additionally, in the second panels from the top, we show the light curve of a semiregular variable from the OGLE collection of LPVs in the Galactic bulge \citep{2013AcA....63...21S}. From the huge OGLE catalog, we selected a typical semiregular variable with a primary period almost identical to that of OGLE-GD-CEP-1884. As the name suggests, semiregular variations show significant phase and amplitude fluctuations, quite distinct from what can be observed in the light curve of OGLE-GD-CEP-1884. While the photometric residuals of our target exhibit some dispersion of points, this light curve is far more stable than what is observed in any LPV with the pulsation period of around 80~d. Due to its modest amplitudes of light changes (about 0.33 mag in the {\it I} band and 0.49 mag in the {\it V} band), OGLE-GD-CEP-1884 cannot be classified as a Mira variable either, as it is generally accepted that the Miras' light curves have amplitudes above 0.8~mag in the {\it I} band and 2.5~mag in the {\it V} band.

As OGLE-GD-CEP-1884 has been ruled out as a pulsating red giant, its apparent red color can only be explained by substantial interstellar reddening in the direction of this object. Given that it is situated in the sky close to the Galactic bulge and nearly at the plane of the Milky Way, the notable extinction toward these regions is not surprising (see Section~\ref{sec:dist}).

Is it possible that OGLE-GD-CEP-1884 represents a different type of pulsating stars than classical Cepheids? Within the family of radial pulsators, similar periods can be achieved by type II~Cepheids of the RV~Tauri subtype. The third panels from the top of Figure~\ref{fig2} present a typical time-series photometry of an RV~Tauri star from the OGLE collection \citep{2017AcA....67..297S}. A distinctive feature of this type of variable stars is the alternating deeper and shallower minima of the light curve, a pattern not observed in OGLE-GD-CEP-1884. Moreover, light curves of RV Tauri stars usually exhibit larger amplitudes and show significant variability from cycle to cycle. The absence of these characteristics eliminates the possibility that OGLE-GD-CEP-1884 is a type~II Cepheid.

The last type of stellar variability that we considered as a potential explanation for OGLE-GD-CEP-1884 is the rotation of a chromospherically active (spotted) star. Theoretically, spots on the surface of a star can produce a light curve that closely mimics the luminosity changes of a pulsating star. However, light curves of spotted stars continuously change their shape as the spot coverage evolves on the stellar surface. This effect is clearly visible in the lower panels of Figure~\ref{fig2} that display the OGLE time-series photometry a typical spotted variable \citep{2019ApJ...879..114I}. The stability of brightness variations observed over a period of more than 10 years unequivocally rules out the possibility that OGLE-GD-CEP-1884 is a spotted star.

The ultimate confirmation of the hypothesis that our object is a classical Cepheid is the characteristic triangular shape of its radial-velocity curve published by \citet{2023A&A...680A..36G}. The lower panel of Figure~\ref{fig1} shows velocity measurements from the {\it Gaia} Focused Product Release with the fitted synthetic radial-velocity curve generated by the \citet{2012ApJ...748..107P} code. To achieve the optimal alignment of the template radial-velocity curve with the observations, we shifted its phase by 0.05 relative to the {\it V}-band model light curves. We attribute this discrepancy to inaccuracies present in both the template and observed radial-velocity and light curves.

Considering the above findings, OGLE-GD-CEP-1884 has been added to the OGLE Collection of Galactic Cepheids\footnote{{\it https://ogle.astrouw.edu.pl $>$ OGLE Collection of Variable Stars}\\{\it https://www.astrouw.edu.pl/ogle/ogle4/OCVS/gd/cep/}} \citep{2018AcA....68..315U,2020AcA....70..101S}. Here, we provide not only the basic parameters of variable stars but also their long-term OGLE light curves, enabling researches to conduct the comprehensive studies on these objects.

\section{Discussion}

\subsection{Distance to OGLE-GD-CEP-1884}\label{sec:dist}

The classical Cepheid with the longest-known pulsation period is expected to be the most luminous one in the Milky Way. To directly determine the absolute magnitude of this object, knowledge of its distance is necessary. Unfortunately, the trigonometric parallax of OGLE-GD-CEP-1884 (Table~\ref{tab:table1}) published by \citet{2021A&A...649A...1G} has a relative uncertainty exceeding 40\%, leading to significant uncertainty in the distance. According to \citet{2021AJ....161..147B}, the geometric distance to OGLE-GD-CEP-1884 falls within the range of 3640--5712~pc (68\% confidence interval), with a median value of 4646~pc.

This result can be compared to the distance determined from the period--luminosity relation obeyed by classical Cepheids. In order to minimize the effect of interstellar extinction, we utilized mid-infrared (mid-IR) observations of OGLE-GD-CEP-1884 and mid-IR period--luminosity relations rather than the optical ones, to calculate the Cepheid distance.

OGLE-GD-CEP-1884 has been observed in the mid-IR domain by the Wide-field Infrared Survey Explorer (WISE, \citealt{2010AJ....140.1868W}) in the W1 [3.4$\,\mu$m], W2 [4.6$\,\mu$m], W3 [12$\,\mu$m], and W4 [22$\,\mu$m] bands. We found one counterpart within $1''$ from the location of the Cepheid in the AllWISE \citep{2014yCat.2328....0C}, unWISE \citep{2019ApJS..240...30S}, and CatWISE \citep{2021ApJS..253....8M} catalogs. The mid-IR light curve of the Cepheid, containing 28 individual observations, is also available from the AllWISE Multiepoch Photometry Database, although all datapoints are clustered at roughly two epochs ($\rm{MJD}\approx55251$ and $\rm{MJD}\approx55433$~days). Given the long pulsation period of OGLE-GD-CEP-1884, these observations should in fact be considered as only two datapoints, therefore we did not include the light curve in further analysis. The AllWISE magnitudes are available in all four WISE bands, while the unWISE  and CatWISE in the W1 and W2 bands only, since they are largely based on observations taken during the post-cryogenic NEOWISE mission \citep{2011ApJ...731...53M}. The saturation limits reported for WISE are approximately 8, 7, 3.8, and -0.4~mag for W1, W2, W3 and W4 bands, respectively. However, the WISE profile-fitting photometry, which for the saturated stars is based on the amount of flux in the wings of the point spread function, is considered to be reliable up to approximately 2.0, 1.5, -3.0, and -4.0 mag in W1, W2, W3, and W4 bands, respectively.

The Cepheid has also been observed by the {\it Spitzer} Space Telescope \citep{2004ApJS..154....1W} in the I1 [3.6$\,\mu$m], I2 [4.5$\,\mu$m], I3 [5.8$\,\mu$m] and I4 [8.0$\,\mu$m] bands as part of the Galactic Legacy Infrared Midplane Survey Extraordinaire (GLIMPSE) program \citep{2003PASP..115..953B}, although only the I3 and I4 magnitudes are available because of the star's large brightness (the reported saturation limits for GLIMPSE observations are 7, 6.5, 4, and 4~mags in I1, I2, I3, and I4 bands, respectively).

All mid-IR brightness measurements of OGLE-GD-CEP-1884 from the WISE and {\it Spitzer} catalogs are listed in Table~\ref{tab:table2}. Our Cepheid is a very bright star in the mid-IR wavelengths, much brighter than the saturation limit in the {\it W}1 and {\it W}2 bands, although still in the magnitude range considered reliable. On the other hand, there is a notable difference, of the order of 0.3~mag, between W1 measuremets in AllWISE, unWISE, and CatWISE. This is most probably caused by different handling of saturated stars in the AllWISE, unWISE, and CatWISE projects. \cite{2019ApJS..240...30S} reported a magnitude offset of $\sim$0.2~mag among different catalogs for stars brighter than $\sim$8~mag in W1 and W2 bands (see their Figure~7). On the other hand, W3 and W4 AllWISE magnitudes, as well as I3 and I4 GLIMPSE magnitudes are below the saturation limits.

We calculated the distance to the Cepheid using the mid-IR period--luminosity relations provided by \cite{2018ApJ...852...78W}, calibrated specifically for the WISE and {\it Spitzer} passbands. The extinction, even though small at the mid-IR wavelengths, is non negligible within the Galactic plane, therefore we used the {\it mwdust} 3D extinction maps that provide extinction values in the $K_S$ band for a given on-sky location and distance \citep{2016ApJ...818..130B}. The $A_{K_{S}}$ extinction value was then transformed to the WISE and {\it Spitzer} bands using the mid-IR extinction curve from \cite{2016ApJS..224...23X}. Since the extinction value is distance-dependent, we calculated the distances iteratively, initially assuming no extinction. Then, we extracted the $A_{K_S}$ value for that distance, transformed it into the mid-IR bands, and recalculated the distance with extinction. The process was repeated until the distance did not change. Distance values in all available passbands are listed in Table~\ref{tab:table2}. The extinction value extracted from {\it mwdust} is $A_{K_{S}}\approx0.84$~mag, and after transforming to mid-IR $A_{\rm{W1}}\approx0.47$, $A_{\rm{W2}}\approx0.40$, $A_{\rm{W3}}\approx0.43$, $A_{\rm{W4}}\approx0.34$, $A_{\rm{I1}}\approx0.40$, and $A_{\rm{I2}}\approx0.39$~mag.

The final Cepheid distance, calculated as the median of all the individual measurements, is $d=4472$~pc. We use the standard deviation of individual distances $\sigma=332$~pc as the distance uncertainty, rather than relying on brightness uncertainties, which are significantly smaller than brightness differences between the WISE catalogs. When we reject all $W1$ and $W2$ mesurements from Table~\ref{tab:table2}, leaving only $W3$, $W4$, I3, and I4 magnitudes, we coincidentally arrive at the same distance value, i.e., $d=4472$~pc, with the standard deviation $\sigma=166$~pc. As can be seen, the distance obtained from the Leavitt law agrees well with the distance derived from the {\it Gaia} parallax, providing additional confirmation that OGLE-GD-CEP-1884 is indeed a classical Cepheid.

The distance to the Cepheid can be coupled with its proper motion and mean radial velocity from {\it Gaia} (Table~\ref{tab:table1}) to measure the 3D velocity vector of the object \citep{2019ApJ...870L..10M}. The velocity component toward the Galactic center is $U = -9.9 \pm 8.5\,\mathrm{km\,s}^{-1}$, in the direction of Galactic rotation $V=220.8 \pm 3.9\,\mathrm{km\,s}^{-1}$, and toward the North Galactic pole $W=-3.4 \pm 1.8\,\mathrm{km\,s}^{-1}$. This velocity vector is consistent with the Galactic rotation curve measured from classical Cepheids \citep{2019ApJ...870L..10M}. It also confirms the classification of the star as a Cepheid -- a velocity vector inconsistent with the rotation curve would indicate incorrect distance and/or classification.

\setlength{\tabcolsep}{7pt}
\begin{deluxetable*}{ccccccccccccc}
\tablecolumns{13}
\tablecaption{Mid-IR magnitudes and distances of OGLE-GD-CEP-1884\label{tab:table2}}
\tablehead{ \colhead{Catalog} & \colhead{W1} & \colhead{$d_{\rm{W1}}$} & \colhead{W2} & \colhead{$d_{\rm{W2}}$} & \colhead{W3} & \colhead{$d_{\rm{W3}}$} & \colhead{W4} & \colhead{$d_{\rm{W4}}$} & \colhead{I3} & \colhead{$d_{\rm{I3}}$} & \colhead{I4} & \colhead{$d_{\rm{I4}}$} \\
\colhead{} & \colhead{[mag]} & \colhead{[pc]} & \colhead{[mag]} & \colhead{[pc]} & \colhead{[mag]} & \colhead{[pc]} & \colhead{[mag]} & \colhead{[pc]} & \colhead{[mag]} & \colhead{[pc]} & \colhead{[mag]} & \colhead{[pc]} }
\startdata
        AllWISE         & 4.96 & 4319 & 4.90 & 4417 & 4.98 & 4461 & 4.68 & 4275 & --   & --   & --   & --   \\
        unWISE          & 5.36 & 5121 & 4.96 & 4543 & --   &  --  &  --  &  --  & --   & --   & --   & --   \\
        CatWISE         & 5.38 & 5161 & 4.72 & 4092 & --   &  --  &  --  &  --  & --   & --   & --   & --   \\
{\it Spitzer} GLIMPSE   & --   & --   & --   & --   & --   &  --  &  --  &  --  & 4.99 & 4483 & 4.91 & 4741 \\
\enddata
\end{deluxetable*}

\begin{figure*}
\centering
\includegraphics[width=16cm]{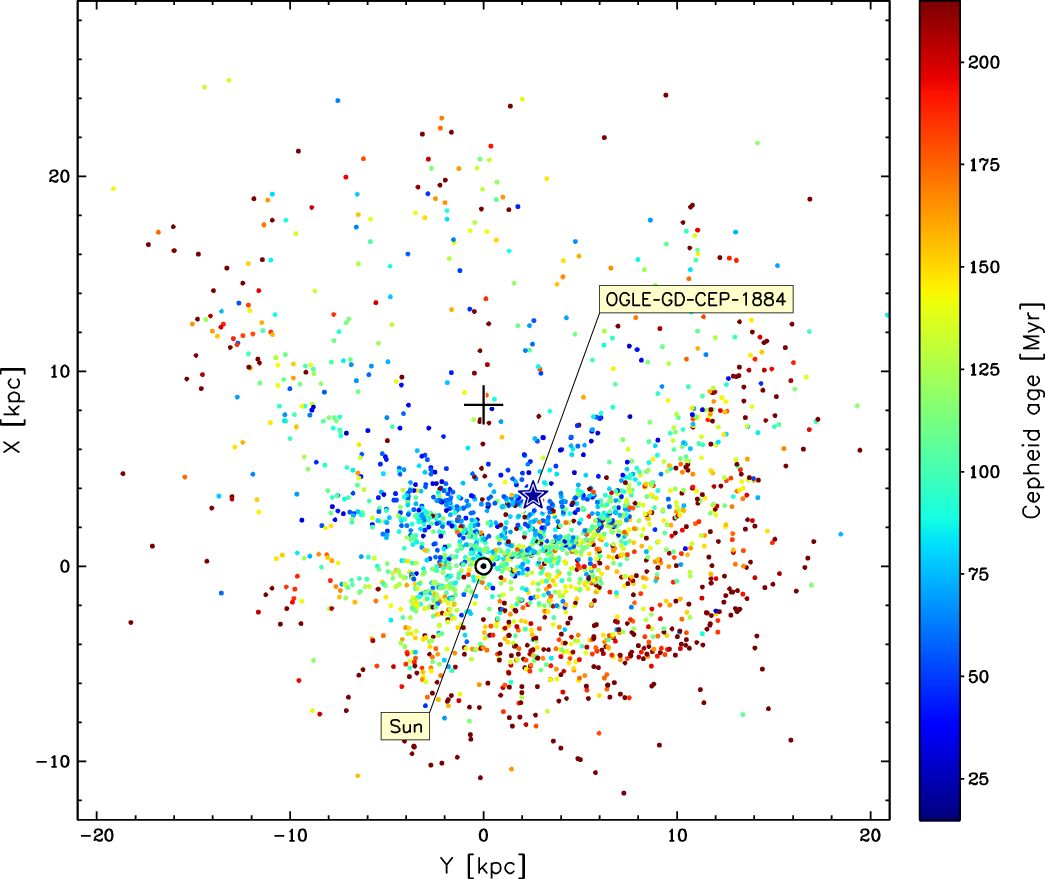}
\caption{Face-on view of the Milky Way traced by 2388 classical Cepheids with distances measured using mid-IR period--luminosity relations \citep{2019Sci...365..478S,2019AcA....69..305S}. The colors of the points indicate ages of the Cepheids (as shown by the scale on the right side) derived with the period--age relation by \citet{2016A&A...591A...8A}. OGLE-GD-CEP-1884 is marked by the star symbol, the Sun is indicated by the white circle, and the Galactic center is shown by the black cross.\label{fig3}}
\end{figure*}

\subsection{OGLE-GD-CEP-1884 on the Map of the Galaxy}\label{sec:ulpc}

Having the distance to OGLE-GD-CEP-1884, we can place it on the Galaxy map along with 2387 other classical Cepheids from \citet{2019Sci...365..478S,2019AcA....69..305S}. Figure~\ref{fig3} presents the top view of the Milky Way traced by Cepheids with the mid-IR measurements obtained from the WISE or {\it Spitzer} missions. Points are color-coded according to the Cepheid ages, estimated using the period--age relations derived by \citet{2016A&A...591A...8A}. We assumed the initial rotation of $\omega=0.5$ and took into account the metallicity gradient in our Galaxy reported by \cite{2014A&A...566A..37G}. Following \citet{2019Sci...365..478S}, we calculated the age of OGLE-GD-CEP-1884 assuming $\omega=0.5$ as well, and an unknown crossing number and location of our Cepheid within the instability strip. The resulting age of OGLE-GD-CEP-1884 is 22~Myr. As noted by \cite{2016A&A...591A...8A}, individual Cepheid ages estimated with the above assumptions are uncertain by up to $\sim$50\%.

Figure~\ref{fig3} shows that OGLE-GD-CEP-1884 aligns well with the general distribution of classical Cepheids in the Milky Way. Our variable is situated in the Norma Arm, in the region predominantly occupied by relatively young Cepheids. Its distance from the Galactic center is approximately 5.3~kpc. It is surprising that the most luminous Cepheid in the Galaxy, located relatively close to us, has not been identified until now.

\subsection{OGLE-GD-CEP-1884 as an Ultra Long Period Cepheid}\label{sec:ulpc}

The pulsation period of OGLE-GD-CEP-1884 falls just below the arbitrary lower limit period of 80~d adopted for the so-called ultra long period (ULP) Cepheids \citep{2009ApJ...695..874B}. In fact, some authors include Cepheids with slightly shorter periods in this group. For example, \citet{2015AJ....149...66N} considered the M31 variable 8-0326 (with a period of 74.43~d) as a candidate for a ULP Cepheid. Applying this relaxed definition, OGLE-GD-CEP-1884 can be recognized as the first ULP Cepheid identified in our Galaxy.

The significance of ULP Cepheids lies in their high absolute brightness, which enables distance measurements to galaxies up to $\sim$100~Mpc \citep[e.g.][]{2012Ap&SS.341..143F}, whereas the range of ordinary (shorter-period) classical Cepheids typically extends only to about 30~Mpc \citep[e.g.][]{2016ApJ...826...56R}. To date, not a single ULP Cepheid has been identified in the Milky Way, which is challenging to explain given the prevalence of such objects in neighboring galaxies. For instance, the Large Magellanic Cloud hosts four ULP Cepheids \citep{1971SCoA...13.....P}, the Small Magellanic Cloud -- three \citep{1966SCoA....9....1P}, M31 -- seven \citep{2015AJ....149...66N,2018AJ....156..130K}, NGC~55 -- five \citep{2006AJ....132.2556P}, while NGC~4038 -- nine \citep{2016ApJ...830...10H,2016ApJ...826...56R}. Cumulatively, a total of 72 ULP Cepheids have been identified within 26 neighboring galaxies so far \citep{2022Univ....8..335M}. The discovery of OGLE-GD-CEP-1884 suggests that ULP Cepheids are also present in our Galaxy, probably in the regions strongly affected by interstellar extinction. This finding underscores the need for current and future sky surveys to adopt more efficient methods of detecting and classifying variable stars.

\section{Conclusions}

We reported the discovery of the longest-period classical Cepheid known in the Galaxy. Its pulsation period, 78.14~d, is nearly 10~d longer than that of the second-longest-period Cepheid, S Vulpeculae. OGLE-GD-CEP-1884 is likely the most luminous, youngest, and most massive Cepheid known in the Milky Way. It can be utilized for improving the fitting of the Leavitt law, investigating the youngest structures of the Galaxy, and studying the evolution of massive stars.

OGLE-GD-CEP-1884 is quite a bright star in the sky, particularly at longer wavelengths, so it is surprising that it has been overlooked in existing catalogs of Cepheids until now. This suggests that there might be more such long-period classical Cepheids in the Galaxy that remain undetected, and the Milky Way does not differ in the number of long-period Cepheids compared to other galaxies. The fact that this star was misclassified as an LPV in the ASAS-SN and {\it Gaia} catalogs indicates the need for continuous improvement of algorithms for automated classification of variable stars. For now, traditional visual inspection of light curves remains the most reliable method for classifying variable stars. 

\begin{acknowledgments}
The paper has been funded by the European Union (ERC, LSP-MIST, 101040160). Views and opinions expressed are however those of the authors only and do not necessarily reflect those of the European Union or the European Research Council. Neither the European Union nor the granting authority can be held responsible for them. This work has been co-funded by the National Science Centre, Poland, grant no.~2022/45/B/ST9/00243. For the purpose of Open Access, the author has applied a CC-BY public copyright license to any Author Accepted Manuscript (AAM) version arising from this submission. PI has been supported by the National Science Center, Poland, grant no.~2019/35/N/ST9/02474.
\end{acknowledgments}


\begin{thebibliography}{}

\bibitem[Anderson et al.(2016)]{2016A&A...591A...8A} Anderson, R.~I., Saio, H., Ekstr{\"o}m, S., et al.\ 2016, \aap, 591, A8. doi:10.1051/0004-6361/201528031
\bibitem[Bailer-Jones et al.(2021)]{2021AJ....161..147B} Bailer-Jones, C.~A.~L., Rybizki, J., Fouesneau, M., et al.\ 2021, \aj, 161, 147. doi:10.3847/1538-3881/abd806
\bibitem[Benjamin et al.(2003)]{2003PASP..115..953B} Benjamin, R.~A., Churchwell, E., Babler, B.~L. et al.\ 2003, \pasp, 115, 953. doi:10.1086/376696 
\bibitem[Bird et al.(2009)]{2009ApJ...695..874B} Bird, J.~C., Stanek, K.~Z., \& Prieto, J.~L.\ 2009, \apj, 695, 874. doi:10.1088/0004-637X/695/2/874
\bibitem[Bovy et al.(2016)]{2016ApJ...818..130B}  Bovy, J., Rix, H.-W., Green, G.~M., et al.\ 2016, \apj, 818, 130. doi:10.3847/0004-637X/818/2/130
\bibitem[Cutri et al.(2014)]{2014yCat.2328....0C} Cutri, R.~M., Wright, E.~L., Conrow, T., et al.\ 2014, VizieR On-line Data Catalog: II/328, 2328, 0.
\bibitem[Fernie(1970)]{1970AJ.....75..244F} Fernie, J.~D.\ 1970, \aj, 75, 244. doi:10.1086/110970
\bibitem[Fiorentino et al.(2012)]{2012Ap&SS.341..143F} Fiorentino, G., Clementini, G., Marconi, M., et al.\ 2012, \apss, 341, 143. doi:10.1007/s10509-012-1043-4
\bibitem[Gaia Collaboration et al.(2021)]{2021A&A...649A...1G} Gaia Collaboration, Brown, A.~G.~A., Vallenari, A., et al.\ 2021, \aap, 649, A1. doi:10.1051/0004-6361/202039657
\bibitem[Gaia Collaboration et al.(2023)]{2023A&A...680A..36G} Gaia Collaboration, Trabucchi, M., Mowlavi, N., et al.\ 2023, \aap, 680, A36. doi:10.1051/0004-6361/202347287
\bibitem[Genovali et al.(2014)]{2014A&A...566A..37G} Genovali, K., Lemasle, B., Bono, G., et al.\ 2014, \aap, 566, 37. doi:10.1051/0004-6361/201323198
\bibitem[Hackstein et al.(2015)]{2015AN....336..590H} Hackstein, M., Fein, C., Haas, M., et al.\ 2015, Astron. Nachr., 336, 590. doi:10.1002/asna.201512195
\bibitem[Hoffmann et al.(2016)]{2016ApJ...830...10H} Hoffmann, S.~L., Macri, L.~M., Riess, A.~G., et al.\ 2016, \apj, 830, 10. doi:10.3847/0004-637X/830/1/10
\bibitem[Iwanek et al.(2019)]{2019ApJ...879..114I} Iwanek, P., Soszy{\'n}ski, I., Skowron, J., et al.\ 2019, \apj, 879, 114. doi:10.3847/1538-4357/ab23f6
\bibitem[Jayasinghe et al.(2019)]{2019MNRAS.486.1907J} Jayasinghe, T., Stanek, K.~Z., Kochanek, C.~S., et al.\ 2019, \mnras, 486, 1907. doi:10.1093/mnras/stz844
\bibitem[Kodric et al.(2018)]{2018AJ....156..130K} Kodric, M., Riffeser, A., Hopp, U., et al.\ 2018, \aj, 156, 130. doi:10.3847/1538-3881/aad40f
\bibitem[Lebzelter et al.(2023)]{2023A&A...674A..15L} Lebzelter, T., Mowlavi, N., Lecoeur-Taibi, I., et al.\ 2023, \aap, 674, A15. doi:10.1051/0004-6361/202244241
\bibitem[Mainzer et al.(2011)]{2011ApJ...731...53M} Mainzer, A., Bauer, J., Grav, T., et al.\ 2011, \apj, 731, 53. doi:10.1088/0004-637X/731/1/53
\bibitem[Marocco et al.(2021)]{2021ApJS..253....8M} Marocco, F., Eisenhardt, P.~R.~M., Fowler, J.~W., et al.\ 2021, \apjs, 253, 8. doi:10.3847/1538-4365/abd805
\bibitem[Mr{\'o}z et al.(2019)]{2019ApJ...870L..10M} Mr{\'o}z, P., Udalski, A., Skowron, D.~M., et al.\ 2019, \apjl, 870, L10. doi:10.3847/2041-8213/aaf73f
\bibitem[Musella(2022)]{2022Univ....8..335M} Musella, I.\ 2022, Universe, 8, 335. doi:10.3390/universe8060335
\bibitem[Ngeow et al.(2015)]{2015AJ....149...66N} Ngeow, C.-C., Lee, C.-H., Yang, M.~T.-C., et al.\ 2015, \aj, 149, 66. doi:10.1088/0004-6256/149/2/66
\bibitem[Payne-Gaposchkin \& Gaposchkin(1966)]{1966SCoA....9....1P} Payne-Gaposchkin, C. \& Gaposchkin, S.\ 1966, Smithsonian Contributions to Astrophysics, 9, 1
\bibitem[Payne-Gaposchkin(1971)]{1971SCoA...13.....P} Payne-Gaposchkin, C.~H.\ 1971, Smithsonian Contributions to Astrophysics, 13
\bibitem[Pejcha \& Kochanek(2012)]{2012ApJ...748..107P} Pejcha, O. \& Kochanek, C.~S.\ 2012, \apj, 748, 107. doi:10.1088/0004-637X/748/2/107
\bibitem[Pietrukowicz et al.(2017)]{2017NatAs...1E.166P} Pietrukowicz, P., Dziembowski, W.~A., Latour, M., et al.\ 2017, Nature Astronomy, 1, 0166. doi:10.1038/s41550-017-0166
\bibitem[Pietrukowicz et al.(2021)]{2021AcA....71..205P} Pietrukowicz, P., Soszy{\'n}ski, I., \& Udalski, A.\ 2021, \actaa, 71, 205. doi:10.32023/0001-5237/71.3.2
\bibitem[Pietrzy{\'n}ski et al.(2006)]{2006AJ....132.2556P} Pietrzy{\'n}ski, G., Gieren, W., Soszy{\'n}ski, I., et al.\ 2006, \aj, 132, 2556. doi:10.1086/508927
\bibitem[Riess et al.(2016)]{2016ApJ...826...56R} Riess, A.~G., Macri, L.~M., Hoffmann, S.~L., et al.\ 2016, \apj, 826, 56. doi:10.3847/0004-637X/826/1/56
\bibitem[Schlafly et al.(2019)]{2019ApJS..240...30S} Schlafly, E.~F., Meisner, A.~M., \& Green, G.~M.\ 2019, \apjs, 240, 30. doi:10.3847/1538-4365/aafbea
\bibitem[Skowron et al.(2019a)]{2019Sci...365..478S} Skowron, D.~M., Skowron, J., Mr{\'o}z, P., et al.\ 2019a, Science, 365, 478. doi:10.1126/science.aau3181
\bibitem[Skowron et al.(2019b)]{2019AcA....69..305S} Skowron, D.~M., Skowron, J., Mr{\'o}z, P., et al.\ 2019b, \actaa, 69, 305. doi:10.32023/0001-5237/69.4.1
\bibitem[Soszy{\'n}ski et al.(2008)]{2008AcA....58..293S} Soszy{\'n}ski, I., Udalski, A., Szyma{\'n}ski, M.~K., et al.\ 2008, \actaa, 58, 293. doi:10.48550/arXiv.0811.3636
\bibitem[Soszy{\'n}ski et al.(2013)]{2013AcA....63...21S} Soszy{\'n}ski, I., Udalski, A., Szyma{\'n}ski, M.~K., et al.\ 2013, \actaa, 63, 21. doi:10.48550/arXiv.1304.2787
\bibitem[Soszy{\'n}ski et al.(2016)]{2016MNRAS.463.1332S} Soszy{\'n}ski, I., Smolec, R., Dziembowski, W.~A., et al.\ 2016, \mnras, 463, 1332. doi:10.1093/mnras/stw1933
\bibitem[Soszy{\'n}ski et al.(2017)]{2017AcA....67..297S} Soszy{\'n}ski, I., Udalski, A., Szyma{\'n}ski, M.~K., et al.\ 2017, \actaa, 67, 297. doi:10.32023/0001-5237/67.4.1
\bibitem[Soszy{\'n}ski et al.(2020)]{2020AcA....70..101S} Soszy{\'n}ski, I., Udalski, A., Szyma{\'n}ski, M.~K., et al.\ 2020, \actaa, 70, 101. doi:10.32023/0001-5237/70.2.2
\bibitem[Udalski et al.(2015)]{2015AcA....65....1U} Udalski, A., Szyma{\'n}ski, M.~K., \& Szyma{\'n}ski, G.\ 2015, \actaa, 65, 1. doi:10.48550/arXiv.1504.05966
\bibitem[Udalski et al.(2018)]{2018AcA....68..315U} Udalski, A., Soszy{\'n}ski, I., Pietrukowicz, P., et al.\ 2018, \actaa, 68, 315. doi:10.32023/0001-5237/68.4.1
\bibitem[Wang et al.(2018)]{2018ApJ...852...78W} Wang, S., Chen, X., de Grijs, R., \& Deng, L.\ 2018, \apj, 852, 78. doi:10.3847/1538-4357/aa9d99
\bibitem[Werner et al.(2004)]{2004ApJS..154....1W} Werner, M.~W., Roellig, T.~L., Low, F.~J., et al.\ 2004, \apjs, 154, 1. doi:10.1086/422992
\bibitem[Wright et al.(2010)]{2010AJ....140.1868W} Wright, E.~L., Eisenhardt, P.~R.~M., Mainzer, A.~K., et al.\ 2010, \aj, 140, 1868. doi:10.1088/0004-6256/140/6/1868
\bibitem[Xue et al.(2016)]{2016ApJS..224...23X} Xue, M., Jiang, B.~W., Gao, J., Liu, J., Wang, S., Li, A.\ 2016, \apjs, 224, 23. doi:10.3847/0067-0049/224/2/23

\end{thebibliography}
\end{document}